\documentclass[aps,prb,twocolumn,amsmath,amssymb,superscriptaddress,floatfix,showpacs]{revtex4}
\usepackage{graphicx}
\usepackage{dcolumn}
\usepackage{bm}
\bibstyle{apsrev.bib}

\newcommand{\be}{\begin{equation}}
\newcommand{\ee}{\end{equation}}
\newcommand{\bea}{\begin{eqnarray}}
\newcommand{\eea}{\end{eqnarray}}

\def\opsimeq{\mathop{\simeq}}

\begin{document}

\title{Disordered Potts model on the diamond hierarchical lattice:\\
Numerically exact treatment in the large-$q$ limit}

\author{Ferenc Igl\'oi}%
 \email{igloi@szfki.hu}
 \affiliation{Research Institute for Solid State Physics and Optics, 
P.O.Box 49, H-1525 Budapest, Hungary}
 \affiliation{Institute of Theoretical Physics,
Szeged University, H-6720 Szeged, Hungary}
 \affiliation{Groupe de Physique Statistique, D\'epartement Physique de la Mati\`ere et 
des Mat\'eriaux,
Institut Jean Lamour, CNRS---Nancy Universit\'e---UPV Metz,
BP 70239, F-54506 Vand\oe uvre l\`es Nancy Cedex, France}
\author{Lo\"{\i}c Turban}
 \email{turban@lpm.uhp-nancy.fr}
 \affiliation{Groupe de Physique Statistique, D\'epartement Physique de la Mati\`ere 
et des Mat\'eriaux,
Institut Jean Lamour, CNRS---Nancy Universit\'e---UPV Metz,
BP 70239, F-54506 Vand\oe uvre l\`es Nancy Cedex, France}

\date{\today}

\begin{abstract}
We consider the critical behavior of the random $q$-state Potts model in the large-$q$ limit with different types of disorder leading to
either the nonfrustrated random ferromagnet regime or the frustrated spin-glass regime. The model is studied on the diamond hierarchical
lattice for which the Migdal-Kadanoff real-space renormalization is exact. It is shown to have a ferromagnetic and
a paramagnetic phase and the phase transition is controlled by four different fixed points. The state of the system is characterized
by the distribution of the interface free energy $P(I)$ which is shown to satisfy different integral equations at the fixed points.
By numerical integration we have obtained the corresponding stable laws of nonlinear combination of random numbers and obtained numerically
exact values for the critical exponents.
\end{abstract}


\maketitle

\section{Introduction}
\label{sec:1}

Despite continuous efforts, several properties of many body systems in the presence of quenched disorder
and frustration are still not well understood. Notoriously difficult systems are spin glasses, in particular, in finite dimensions and with finite-range interactions.\cite{review_sg} One basic problem of our understanding in this
field of research is the lack of exact solutions for nontrivial models. Exceptions in this respect
are such systems in which disorder fluctuations are fully dominant and the properties
of the system are governed by an infinite-disorder fixed point.\cite{igloi_monthus} This happens, among others for random quantum 
spin chains and for a class of stochastic models with quenched disorder. If, however, the properties of the
system are controlled by a conventional (finite-disorder) random fixed point, such as for classical
spin glasses, the exact results are scarce.

In these cases, in order to obtain more accurate information,
one often considers hierarchical lattices~\cite{hier_lat} in which the Migdal-Kadanoff renormalization~\cite{migdal_kadanoff} can be performed
exactly. For simple models, such as for directed polymers, one can notice a simple, although nonlinear relation
between the original and the transformed random energies and one can derive closed integral equations between
the corresponding distribution functions, which can then be studied by various methods.\cite{dp,dp1} However for spin models, such as the random-bond Ising model, it is generally not possible to write renormalization equations directly for the distribution functions.
In these cases one can treat numerically large finite samples exactly and average the obtained results over quenched
disorder.\cite{young,jayaprakash,mckay,gardner,bray,hilhorst} Although bringing very accurate numerical results, this type of treatment is still not exact and, as usually happens in random systems, the source of inaccuracy comes from the averaging process over quenched
disorder. In such calculations one cannot, for example, decide about the universality of the fixed points, i.e.,
if the critical singularities are independent or not of the form of the initial distribution of the disorder.

In this paper we consider such a spin model, the $q$-state random Potts model~\cite{wu} in the large-$q$ limit, for which
the Migdal-Kadanoff renormalization leads to closed integral equations in terms of the distribution functions. In this
respect our results are comparable with those obtained for directed polymers on hierarchical lattices.\cite{dp}
The $q$-state Potts model, as it is well known, is equivalent for $q=2$ to the Ising model.
On a hypercubic lattice for sufficiently large $q$ the transition
turns to first order, whereas on a hierarchical lattice the transition for any finite $q$
stays of second order, but the critical exponents are $q$ dependent.\cite{derrida}

Properties of the Potts model with random ferromagnetic and antiferromagnetic couplings ($\pm J$ model)
have been studied numerically to some extent. On the cubic lattice a spin-glass (SG) phase has been identified
for $q=3$ (Ref.~15) but the SG phase is absent for large enough value of $q$, as observed for $q=10$.\cite{binder} On the square lattice,
the dimension being below the critical one, $d_c>2$,
there is no SG phase, and the phase diagram consists only of the ferromagnetic and the paramagnetic
phases. The transition between these phases has been studied numerically for $q=3$ where it is
found to be controlled by four different fixed points\cite{huse} (see also in Ref.\cite{picco}).The model has been also considered on the diamond
hierarchical lattice with random ferromagnetic couplings.\cite{kinzel,derrida_gardner,andelman,monthus_garel}

Here we extend these studies by including frustration, too, and investigate the large-$q$ limit. We note
that in the disordered case the large-$q$ limit is generally not singular,\cite{cardy,bertrand} the critical behavior is qualitatively
similar to that of finite $q$ and also the critical exponents are smoothly saturating as $q \to \infty$.\cite{qlarge}
We mention that it has been recently suggested that, in the large-$q$ limit,   the Potts model is a plausible model for supercooled liquids. This is indeed the case within the mean-field approach~\cite{potts_liquid} but it seems to be
completely different for the nearest-neighbor model.\cite{binder,lee}

To consider the large-$q$ limit of the model leads to technical simplifications, at least for random
ferromagnetic couplings. The high-temperature series expansion of the model is dominated by a single
diagram, the properties of which have been studied in $2d$ and $3d$ by combinatorial optimization  
methods.\cite{qlarge} This type of simplification is valid for the hierarchical lattice, too, but
for this lattice the exact renormalization holds for antiferromagnetic couplings, too. As we will
see the renormalized parameters in this limit are expressed as simple but nonlinear combinations of
the original parameters, such as the interface free energy. This makes it possible to write integral
equations for the distribution functions which are studied by various methods.

The structure of the paper is the following: the $q$-state Potts model and its Migdal-Kadanoff
renormalization in the large-$q$ limit is presented in Sec.~\ref{sec:2}. The phase diagram of the
model is calculated by the numerical pool method in Sec.~\ref{sec:3}, whereas the properties of the fixed points
are obtained through numerical integration in Sec.~\ref{sec:4}. Our results are discussed in Sec.~\ref{sec:5}
and details of the derivation of the integral equations at the fixed points are given in the Appendices.

\section{Model and renormalization}
\label{sec:2}


\begin{figure}
\vglue 5mm
\begin{center}
\includegraphics[width=6.cm,angle=0]{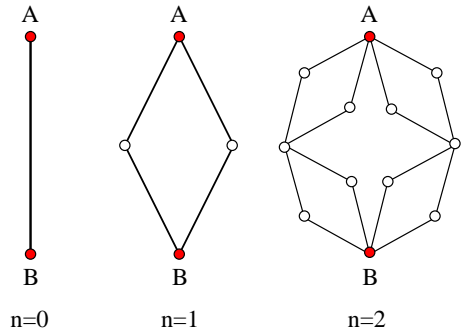}
\end{center}
\caption{(Color online) First steps of the construction of the hierarchical diamond lattice with $b=2$.}
\label{fig1}
\end{figure}

We consider the $q$-state Potts model defined by the Hamiltonian:
\begin{equation}
{\cal H}=-\sum_{\langle i,j \rangle} J_{ij} \delta(\sigma_i,\sigma_j)
\label{hamilton}
\end{equation}
in terms of the Potts-variables  $\sigma_i=1,2,\dots,q$, associated with the sites indexed by $i$ of the lattice.
Here the summation runs over nearest-neighbor pairs and the couplings $J_{ij}$ are independent and identically
distributed random numbers, which can be either positive or negative. We consider the large-$q$ limit, in which
case it is convenient to rescale the temperature~\cite{qlarge} as $T'=T\ln q$, so that $e^{\beta}=q^{\beta'}$ where $\beta'=1/T'$ ($k_B=1$) and consider
the high-temperature series expansion of the model, in which the partition function ${\cal Z}$ is dominated by one
diagram:
\begin{equation}
{\cal Z} \opsimeq_{q \to \infty} q^{\phi} + {\rm subleading~terms}
\label{dominant}
\end{equation}
which is related to the free energy through $\phi=-\beta' F$.

In this paper the Potts model is considered on the diamond hierarchical lattice,
which is constructed recursively
from a single link corresponding to the generation $n=0$.
(see Fig.~\ref{fig1}).
The generation $n=1$ consists of $b$ branches in parallel, each branch
 containing two bonds in series.
 The next generation $n=2$ is obtained by applying the same transformation
to each bond of the generation $n=1$.
At generation $n$, the length $L_n$ measured by the number of bonds between the two extreme sites
$A$ and $B$ is $L_n=2^n$, and the total number of bonds is 
\begin{eqnarray}
B_n=(2b)^n = L_n^{d_{eff}(b)} \ \ \ {\rm \ \ with \  \ } 
d_{eff}(b)= \frac{ \ln (2b)}{\ln 2}
\label{bn}
\end{eqnarray}
where $d_{eff}(b)$ represents some effective dimensionality.

In the following we consider two different boundary conditions and denote the
partition function as ${\cal Z}^{1,1}_n$ and ${\cal Z}^{1,2}_n$, when the two extreme sites $A$ and $B$
are fixed in the same state or in different states, respectively. Their ratio
is given by:
\begin{equation}
x_n=\frac{{\cal Z}^{1,1}_n}{{\cal Z}^{1,2}_n}=q^{\beta' F^{\rm inter}_n}\;,
\label{x_n}
\end{equation}
where $F^{\rm inter}_n=F^{1,2}_n-F^{1,1}_n$ is the interface free energy. The ratio $x_n$
obeys the recursion equation~\cite{kinzel,andelman}
\begin{equation}
x_{n+1}=\prod_{i=1}^b \left[ \frac{x_n^{(i_1)}x_n^{(i_2)}+(q-1)}{x_n^{(i_1)}+x_n^{(i_2)}+(q-2)} \right]
\label{x_recursion}
\end{equation}
Now we consider the large-$q$ limit and keeping in mind that the partition function is given by one dominant term,
see Eq.~(\ref{dominant}), we obtain the following recursion for the scaled interface free energy,
$I_n=\beta' F^{\rm inter}_n$:
\begin{equation}
I_{n+1}=\sum_{i=1}^b \Phi\left[I_{n}^{(i_1)},I_{n}^{(i_2)}\right]\;.
\label{I_recursion}
\end{equation}
Here the auxiliary function is:
\begin{widetext}
\begin{equation}
\Phi\left[I^{(1)},I^{(2)}\right]=
\begin{cases}
0& \text{if $I_{max}+I_{min}<1$ and $I_{max}<1$},\\
1-I_{max}& \text{if $I_{max}+I_{min}<1$ and $I_{max}>1$},\\
I_{max}+I_{min}-1& \text{if $I_{max}+I_{min}>1$ and $I_{max}<1$},\\
I_{min}& \text{if $I_{max}+I_{min}>1$ and $I_{max}>1$}.
\end{cases}
\label{I_recursion1}
\end{equation}
\end{widetext}
with $I_{max}=max(I^{(1)},I^{(2)})$ and $I_{min}=min(I^{(1)},I^{(2)})$. The initial condition is given
by
\begin{equation}
I_0^{(i)}=\beta' J_i
\end{equation}
where $J_i$ is the value of the $i$th coupling.

Note that in the large-$q$ limit the recursion relation in Eq.~(\ref{x_recursion}) are simplified and 
there is now a direct relation in Eqs.~(\ref{I_recursion}) and (\ref{I_recursion1}) between the scaled interface
free energies. This recursion relation involves a somewhat complicated nonlinear combination of random variables and we are interested
in the stable law for their distribution function.

\section{Phase diagram}
\label{sec:3}

\subsection{Non-random system}

In the pure case the recursion relation simplifies:
\begin{equation}
I_{n+1}=
\begin{cases}
0& \text{if $0<I_n \le 1/2$},\\
b(2I_n-1)& \text{if $1/2<I_n \le 1$},\\
bI_n& \text{if $I_n>1$}.
\end{cases}
\label{pure}
\end{equation}
It has a fixed-point at $I_c=b/(2b-1)$. The phase transition at this point is of first order. The dominant
diagram for $I<I_c$ consists of isolated points, whereas for $I>I_c$ the dominant diagram is fully connected
and there is a phase coexistence at $I=I_c$.

\subsection{Random system: Numerical study}

For the random case we consider a diamond lattice with branching number $b=2$, i.e., with an effective dimension
$d_{eff}=2$. First we perform a numerical investigation using a continuous box-like distribution of the couplings
\begin{equation}
{\cal P}(J)=
\begin{cases}
1 & \text{if $\frac{p}{1-p}<J<\frac{1}{1-p}$},\\
0& \text{otherwise}.
\end{cases}
\label{distr_SG}
\end{equation}
with $p \le 1$ and such that the mean value of the couplings is given by:
\begin{equation}
\overline{J}=\frac{1}{2}\left(\frac{1+p}{1-p}\right)\,.
\label{mvJ}
\end{equation}
For $p>0$ all couplings are random ferromagnetic and in the limit $p \to 1$ we have the pure system.
For $p<0$ there are also negative bonds, their fraction is increasing with decreasing $p$.
For $p=-1$ the distribution is symmetric with zero mean and standard deviation $1/\sqrt{12}$. In the numerical calculations
we have used the so-called pool method. Starting with $N$ random variables taken from the original distribution in
Eq.~(\ref{distr_SG}),  we generate a new set of $N$ variables through renormalization at a fixed temperature, $T'$,
using Eqs.~(\ref{I_recursion}) and (\ref{I_recursion1}). These are the elements of the pool at the first generation, which are then used as input for 
the next renormalization step.
We check the properties of the pool at each renormalization by calculating the distribution of the scaled interface free energy, its average
and its variance. In practice we have used a pool of $N=5 \times 10^6$ elements and we went up to $n \sim 70-80$ iterations.

\subsubsection{Phases}

At a given point of the phase diagram, $(p,T')$, the renormalized parameters display two different behaviors, which are governed by two trivial fixed points.
In the paramagnetic phase the scaled interface free energy renormalizes to zero, whereas in the ferromagnetic phase it goes to infinity. We note that in the
large-$q$ limit there is no zero-temperature spin-glass phase, contrary to the known results for $q=2$ and $q=3$.\cite{binder1,huse} We analyze the
properties of these trivial fixed points in details in the following section. 

The two phases
are separated by a phase transition line $T'_c(p)$ which can be calculated accurately for a given pool and its true value can be obtained by averaging over
different pools and taking the $N \to \infty$ limit. The phase diagram is shown in Fig.~\ref{fig2} in the plane $T'/\overline{J}$ versus $p$.  In the phase diagram one may notice a reentrance in the regime $p<0$.  At low temperatures
the system is disordered due to frustration. In the intermediate temperature regime there is an order-through-disorder phenomenon and at high temperature
the system is again disordered  due to thermal fluctuations. Similar features has been analysed before for the $q=3$ model.\cite{gingras}

\subsubsection{Phase-transition lines}

\begin{figure}
\begin{center}
\includegraphics[width=6.cm,angle=270]{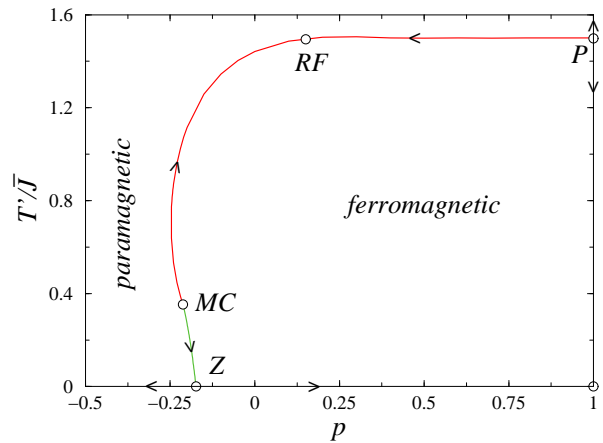}
\end{center}
\caption{(Color online) Phase diagram of the disordered Potts model in the large-$q$ limit on the diamond hierarchical lattice with $b=2$ using the
continuous box-like distribution of Eq.~(\ref{distr_SG}). The critical behavior along the transition line between the paramagnetic and the ferromagnetic phases is controlled by four different fixed points: i) At
$p=1$ the pure systems fixed point (P), ii) from $p=1$ to $p=p_{\rm MC}$ (red line), the random-ferromagnet fixed point (RF), iii) at $p=p_{\rm MC}$ the multicritical fixed point
($\rm MC$), iv) for $p_{\rm MC} < p \le p_{\rm Z}$ (green line)
the zero temperature fixed point (Z).  Note that there is a reentrance
in the phase diagram.}
\label{fig2}
\end{figure}

The properties of the phase transition are different when the disordering effect is dominantly of thermal origin (which happens in the high-temperature part of the
transition line) or when it is dominantly due to frustration (which happens in the low-temperature part of the transition line). At the boundary of the ferromagnetic part of the phase diagram, the pure-system fixed point at $p=1$ is unstable in the presence of any amount of (ferromagnetic) disorder and the transition is controlled
by a new fixed point, the random-ferromagnet (RF) fixed point. According to the phase diagram in Fig.~\ref{fig2}, this fixed point controls the phase transition even
in a part of the region where $p<0$, up to a point $(p_{\rm MC},T'_c(p_{\rm MC}))$. Our numerical studies indicate that at the RF fixed point the scaled interface free
energy is typically  $I_n=O(1)$. Below this temperature, $T'<T'_c(p_{\rm MC})$, the phase transition is controlled by a
zero-temperature (Z) fixed point, the properties of which are very similar to that in the $q=3$ model\cite{huse}, in which it describes the transition between the zero temperature
spin-glass phase and the paramagnetic phase. Our numerical studies show that
along the green line of Fig.~\ref{fig2} between MC and Z, as well as at the Z fixed point the scaled interface free energy grows with the size, $L$, as
\begin{equation}
I(L)=L^{\theta_{\rm Z}} u_I\;,
\label{Ibetac2}
\end{equation}
where the $u_I$ are $O(1)$ random numbers and the droplet exponent is $\theta_{\rm Z}>0$.
Finally at $(p_{\rm MC},T'_c(p_{\rm MC}))$ there is a multicritical (MC) fixed point, analogous to the Nishimori MC point in gauge-invariant systems.\cite{Nishimori}
On the phase diagram shown in Fig.~\ref{fig2}, the coordinates of the fixed points are $(-0.212334;~0.353194)$ for the MC fixed point and  $(-0.173626;~0.0)$
for the Z fixed point, but we did not study the actual position of the RF fixed point.
In the following section we study the properties of the fixed points (except the MC point) through
numerical solution of integral equations.


\section{Properties of the fixed points}
\label{sec:4}

As shown in the numerical study of the previous section, the system has two phases which are controlled by two trivial fixed points and the phase transition line is controlled by four different nontrivial fixed points. In each fixed point there is a characteristic distribution of the scaled interface free-energy parameters, $P(I)$, which transforms under recursion in Eqs.~(\ref{I_recursion}) and (\ref{I_recursion1})
into $P'(I)$. The transformation law of the distribution function can be written in an explicit form, due to the fact that the transformed variables, $I_{n+1}$,
are expressed as a sum of a few $I_n$, which are independent random numbers distributed according to $P(I)$. Thus we are looking for the stable law of
a nonlinear combination of random numbers, which is different in the different fixed points.

\subsection{Paramagnetic phase}

The paramagnetic phase has a trivial limit distribution $P_{para}=\delta(I)$ since all parameters renormalize to zero.

\subsection{Ferromagnetic phase}

In the ferromagnetic phase $I_n$ grows without limit, thus asymptotically for each bond $I_{max} \ge I_{min} > 1$, and in Eqs.~(\ref{I_recursion1})
the last equation holds. Consequently for $b=2$ we have:
\begin{equation}
I_{n+1}=I_n^{(1)}(min)+I_n^{(2)}(min)\;,
\label{I_recursion2}
\end{equation}
which leads the following relation for the distribution function:
\begin{equation}
P'(I)=\int_{0}^{I}\!\! d x\, P_2(x) P_2(I-x)\;,
\label{eq_ferro}
\end{equation}
in terms of
\begin{equation}
P_2(I)=2 P(I) \int_{I}^{\infty}\!\!d x\, P(x) ,\quad I>1\;.
\label{P_2}
\end{equation}
Writing $I$ as $\overline{I}+I_1$ where $\overline{I}$ is the average value and $I_1$ the fluctuating part, Eq.~(\ref{eq_ferro}) leads to a renormalization relation for the average values,  $(\overline{I})'=2\overline{I}$.
Defining the probability distributions for the fluctuating part as $\tilde{P}(I_1)=P(\overline{I}+I_1)$ and $\tilde{P}'(I_1)=P'(2\overline{I}+I_1)$, they  satisfy the following relation:
\begin{eqnarray}
\tilde{P}'(I_1)&=&4 \int_{-\infty}^{\infty}\!\!d x_1\,  \tilde{P}(x_1) \tilde{P}(I_1-x_1) \cr
&\times& \int_{x_1}^{\infty}\!\!d y_1\,\tilde{P}(y_1) \int_{I_1-x_1}^{\infty}\!\!d z_1\,
\tilde{P}(z_1)\;.
\label{eq_ferro'}
\end{eqnarray}
We have studied this equation numerically and found that the fixed-point solution satisfies the relation
\begin{equation}
\tilde{P}'(I_1)=\frac{1}{\lambda}\tilde{P}(I_1/\lambda)\;,
\label{P_ferro}
\end{equation}
with $\lambda=1.230091(1)$. Consequently in the ferromagnetic fixed point the average value and the standard deviation of the scaled
interface free energy are related to the size $L$ of the system by:
\begin{equation}
\overline{I} \sim L^{d_s},\quad \Delta I \sim L^{\theta}\;.
\label{scal_ferro}
\end{equation}
Here the dimension of the interface is $d_s=d_{eff}-1=1$ and the droplet exponent is  $\theta=\log( \lambda)/\log(2)=0.298765(1)$.
Note that the value of the droplet exponent is in agreement with previous numerical studies for finite value of $q$.\cite{monthus_garel} It is also identical to the value
obtained for directed polymers on the same hierarchical lattice.\cite{dp,dp1}

\subsection{The RF fixed point}


\begin{figure}
\begin{center}
\includegraphics[width=6.cm,angle=270]{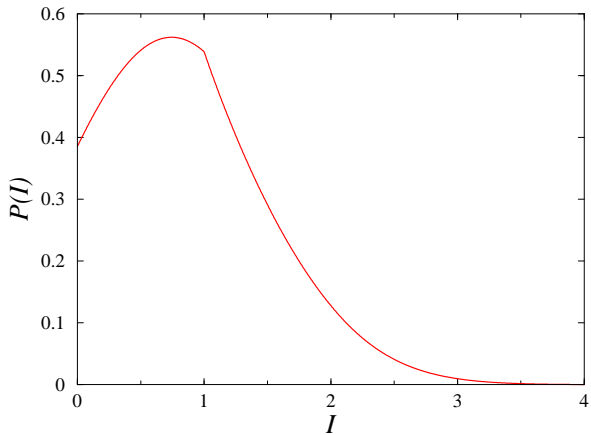}
\end{center}
\caption{(Color online) Probability distribution of the scaled interface free energy at the RF fixed point. At $I=0$ there is a delta peak with strength,
$p_0=0.1280795$, the function is nonanalytic at $I=1$ and $I=2$.}
\label{fig3}
\end{figure}


At the RF fixed point the iterated interface free energy scales to the region $I_n \ge 0$, therefore in the recursion relations of Eq.~(\ref{I_recursion1})
the second is irrelevant. The probability distribution $P(I)$ satisfies different recursion relations at $I=0$ and in
the regions $0<I<1$, $1<I<2$, and $I>2$. These relations are obtained in Appendix \ref{app:1}. At the fixed point the distribution
stays invariant, thus $P'(I)=P(I)$. The fixed point distribution obtained by numerical integration is shown in Fig.~\ref{fig3}. Note that $I$ vanishes with probability $p_0=0.1280795$ and the probability distribution is continuous, but nonanalytic, at $I=1$ and $I=2$.

In order to calculate the thermal eigenvalue of the fixed-point transformation we form the Jacobian $J(x,y)$ through functional
derivation at the fixed point, $J(x,y)=\delta P'(x) / \delta P(y)$, and solve the eigenvalue problem
\begin{equation}
\int \!\!d y\, J(x,y) f_i(y) = \lambda_i^{\rm RF} f_i(x)\;.
\label{eigen}
\end{equation}
In this way we have obtained the leading eigenvalue $\lambda_1^{\rm RF}=1.6994583(1)$, and checked that $\lambda_i^{\rm RF}<1$ for $i \ge 2$, e.g., the correction-to-scaling eigenvalue is 
$\lambda_2^{\rm RF}=0.6378796(1)$. 

The thermal eigenvalue of the transformation is given by:
\begin{equation}
y_t^{\rm RF}=\frac{\log \lambda_1^{\rm RF}}{\log 2}=0.7650750(1)\;,
\label{y_t}
\end{equation}
from which we deduce the correlation length critical exponent  $\nu_{\rm RF}=1/y_t^{\rm RF}=1.307061(1)$. This exponent appears in the scaling form of the average
interface free energy $\overline{I}(T',L)$ in the ordered phase  for $T'<T'_c$ along the transition line in Fig.~\ref{fig2}. We have
\begin{equation}
\overline{I}(T',L)=\left[\frac{L}{\xi_{av}(T')}\right]^{d_s}+\dots\;,
\label{IL}
\end{equation}
where $d_s=d_{eff}-1=1$ and the correlation length diverges as $\xi_{av}(T') \sim (T'-T'_c)^{-\nu_{\rm RF}}$.
Note that the 
fluctuation of the interface free energy $\Delta I(T',L)$ grows with the droplet exponent $\theta$ (see Eq.~(\ref{scal_ferro})) as:
\begin{eqnarray}
\Delta I(T',L)=\left[\frac{L}{\xi_{var}(T')}\right]^{\theta}+\dots\;.
\label{DIL}
\end{eqnarray}
The associated correlation length $\xi_{var}(T')$ is proportional to $\xi_{av}(T')$, thus there is only one
length scale in the problem. The situation seems to be different for finite values of $q$, in which case two
different correlation length exponents are obtained for the average and the fluctuating part of the interface free energy,
respectively.\cite{monthus_garel}

\subsection{The zero temperature fixed point}

The Z fixed point is at zero temperature and here the scaled interface free energy $I_n$ grows to infinity. Therefore
we use the reduced variable $i_n \equiv I_n/{\overline I}_n$ in terms of which the renormalization group
equations in Eq.~(\ref{I_recursion}) are modified as:
\begin{equation}
\frac{I_{n+1}}{{\overline I}_n}=i_{n+1} \alpha_{n+1}=\sum_{i=1}^b \phi\left[i_{n}^{(i_1)},i_{n}^{(i_2)}\right]\;,
\label{i_recursion}
\end{equation}
with $\alpha_{n+1}={\overline I}_{n+1}/{\overline I}_{n}$ and
\begin{widetext}
\begin{equation}
\phi\left[i^{(1)},i^{(2)}\right]=
\begin{cases}
0& \text{if $i_{max}<0$},\\
-i_{max}& \text{if $i_{max}+i_{min}<0$ and $i_{max}>0$},\\
i_{min}& \text{if $i_{max}+i_{min}>0$}.
\end{cases}
\label{i_recursion1}
\end{equation}
\end{widetext}

In agreement with the phase-diagram in Fig.~\ref{fig2} these equations have two trivial fixed
points corresponding to the ferromagnetic phase when $p > p_{\rm Z}$ with $\alpha_{n} \to b$ and to the
paramagnetic phase for $p < p_{\rm Z}$, in which case $\alpha_{n} \to 0$. The nontrivial fixed
point at $p=p_{\rm Z}$ where $\alpha_{n} \to \alpha_{\rm Z}$ governs the zero temperature transition. 

\begin{figure}
\begin{center}
\includegraphics[width=6.cm,angle=270]{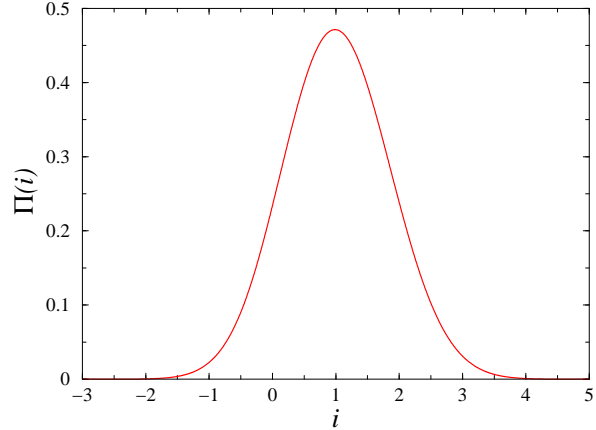}
\end{center}
\caption{(Color online) Probability distribution of the scaled interface free energy, $i=I/\overline{I}$, at the \rm Z fixed point, as well as along
the green line of Fig.~\ref{fig2} between MC and Z. At $i=0$ there is a delta peak with strength,
$\pi_0=0.000158$, the function is nonanalytic at $i=0$.}
\label{fig4}
\end{figure}

The probability
distribution $\Pi(i)$ satisfies different relations for $i<0$, $i>0$, as well as at $i=0$. The corresponding integral equations are presented in Appendix \ref{app:2}.
We have integrated these equations numerically and found that, at the Z fixed point, the probability distribution $\Pi(i)$
transforms as
\begin{equation}
\Pi'(i)=\alpha_{\rm Z}\Pi(i)\;, 
\label{trans-pi}
\end{equation}
with $\alpha_{\rm Z}=1.10661(1)$. Consequently, the droplet
exponent at the zero temperature transition (see Eq.(\ref{Ibetac2})) is given by:
\begin{equation}
\theta_{\rm Z}=\frac{\log \alpha_{\rm Z}}{\log 2}=0.14615(1)\;.
\label{theta_Z}
\end{equation}
The probability distribution at the fixed point is shown in Fig.~\ref{fig4}.

We have calculated the Jacobian at this fixed point, too. Like in Eq.~(\ref{eigen}), from the corresponding  eigenvalue problem we have deduced
the leading eigenvalue $\lambda_1^{\rm Z}=1.49314(1)$ which gives the thermal exponent
\begin{equation}
y_t^{\rm Z}=.57835(1)
\end{equation}
and the correlation-length exponent $\nu_{\rm Z}=1/y_t^{\rm Z}=1.72906(1)$. 

In the ferromagnetic phase the average interface free energy scales along the green transition line in Fig.~\ref{fig2} as
\begin{equation}
\overline{I}(T',L)=\frac{L^{d_s}}{[\xi_{av}(T')]^{d_s-\theta_{\rm Z}}}+\dots\;,
\label{IL1}
\end{equation}
with $d_s=d_{eff}-1=1$. Similarly, the fluctuation of the interface free energy has the scaling form:
\begin{equation}
\Delta I(T',L)=\frac{L^{\theta}}{[\xi_{var}(T')]^{\theta-\theta_{\rm Z}}}+\dots\;.
\label{DIL1}
\end{equation}
Note that these scaling relations differ from those at the RF fixed point; see  Eqs.~(\ref{IL}) and (\ref{DIL}), respectively. The finite-size behavior of these modified forms with $\xi(L)\sim L$ is in agreement with Eq.~(\ref{Ibetac2}).
The correlation lengths $\xi_{av}$ and $\xi_{var}$ diverge as $T' \to T'_c$ with the same critical exponent  $\nu^{\rm Z}$.


\subsection{The MC fixed point}

\begin{figure}
\begin{center}
\includegraphics[width=6.cm,angle=270]{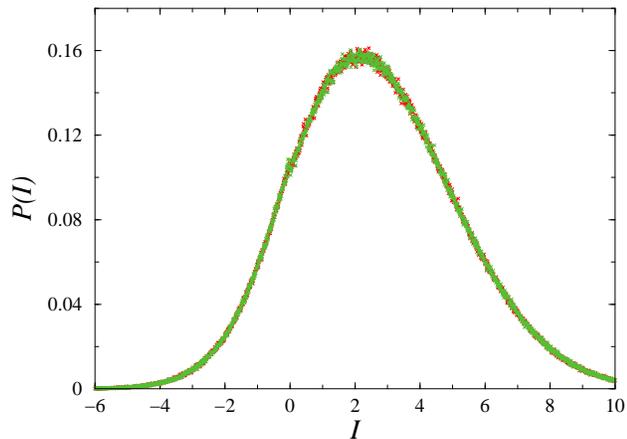}
\end{center}
\caption{(Color online) Distribution of the interface free energy at the
MC point as calculated by the numerical pool method. The green (red) symbols are for $n=12$ ($n=10$)
iterations. A small fraction $n_0 \approx 0.0057$ of the samples have zero interface free energy.}
\label{fig5}
\end{figure}

To complete our study we list here the properties of the MC fixed point, which has been obtained numerically using the pool method. At the MC point, the interface free energy is found to be
an $O(1)$ random number, the distribution of which is shown in Fig.~\ref{fig5}. Both negative and positive couplings are involved in the distribution and the average
value of the interface free energy as well as its fluctuations are larger than at the RF fixed point. In the vicinity of the MC point, the scaling of the interface free energy takes
the form given in Eqs.~(\ref{IL}) and (\ref{DIL}). The divergence of the correlation length can be analyzed  assuming the presence of a multiplicative
logarithmic correction. The numerical data seem to be consistent with the scaling combination
\begin{equation}
\xi(t) \sim (t \ln^{\kappa} t)^{-\nu_{\rm MC}}\;,
\label{xi_tri}
\end{equation}
with $\kappa \approx 1.5$ and $\nu_{\rm MC}=3.61$ for the average as well as for the standard deviation.

\section{Discussion}
\label{sec:5}
We have studied the random Potts model in the large-$q$ limit with such type of disorder which
includes both the random (nonfrustrated) ferromagnet as well as the frustrated spin-glass regime.
The model is considered on the diamond hierarchical lattice, on which the Migdal-Kadanoff
renormalization is exact. First we used the numerical pool method to determine the phase diagram. It consists of a paramagnetic phase and
a ferromagnetic phase, separated by a transition line which is controlled by four fixed points.
This structure of the phase diagram is very similar to that found numerically for the $\pm J$ three-state Potts model in two dimensions,\cite{huse}
although in our model the zero-temperature spin-glass phase is absent.

The state of this random system is shown to be
uniquely determined by the distribution of the scaled interface free-energy, $P(I)$. The renormalization group
transformation is written in the form of integral equations for $P(I)$ and we have studied the properties of its fixed points
by numerical integration.
Mathematically the above problem is equivalent to find the stable law of nonlinear combination of random numbers. 

In the random
ferromagnetic phase the probability distribution is analogous to that of directed polymers.\cite{dp,dp1} For the lattice with $b=2$
the droplet exponent is exactly the same in the two cases. The nontrivial fixed points governing the properties of
the phase transition have different scaling properties, which are summarized in Table~\ref{table1}.

\begin{table}
\caption{Scaling behavior of the average interface free energy, $\overline{I}$, and its fluctuations $\Delta I$, as a function
of the linear size, $L$, in the ferromagnetic and paramagnetic phases and at the different fixed points: P (pure system);
RF (red line between P and MC in Fig.\ref{fig2}); MC; Z (green line between MC and Z in Fig.\ref{fig2}).
The values of the correlation-length critical exponent are also indicated. At the P fixed point the transition is of first order.\label{table1}}
\begin{ruledtabular}
\begin{tabular}{lccc}
& $\overline{I}$ & $\Delta I$ & $\nu$ \\
\hline
$\rm Ferro$ & $ L^{d_s}$ & $L^{\theta}$ &  \\
$\rm Para$ & $0$ & $0$ &  \\
\hline
$\rm P$ & $b/(2b-1)$ & $0$ & $1/d$ \\
$\rm RF$ & $O(1)$ & $O(1)$ & $1.31$ \\
$\rm MC$ & $O(1)$ & $O(1)$ & $3.61$ \\
$\rm Z$ & $L^{\theta_{\rm Z}}$ & $L^{\theta_{\rm Z}}$ & $1.73$ \\
\end{tabular}
\end{ruledtabular}
\end{table}

We note that at the $RF$ fixed point the correlation-length exponent $\nu_{\rm RF}$ corresponds to a mostly thermal scaling field, whereas
at the Z\rm Zfixed point $\nu_{\rm Z }$ is mainly due to a disorder-like scaling field.
In the large-$q$ limit, the same correlation length exponent governs the scaling form of both the average and the fluctuation part of the interface free energy.

Our results are obtained strictly in the large-$q$ limit and our analysis is mainly done for a diamond hierarchical lattice
with $b=2$. Here we comment on possible extensions of our study in three directions. i) For finite, but not too large
values of $q$ the phase diagram is similar to ours, for example the SG phase appears only for $q$-values which are not too far from two. The critical exponents are
expected to have corrections in powers of $1/q$. ii) One can repeat the calculation for larger values of $b$, which corresponds
to a larger effective dimensionality. However the integral equations for the probability distribution $P(I)$ become more
and more complicated: $P(I)$ is a piece-wise function with more and more different intervals of definition. iii) Finally, the calculation can be extended
to other quantities such as the interface energy and the magnetization. For these quantities, however, one should work
with conditional probabilities, which will result in even more complicated equations.

\begin{acknowledgments}
F.I. is grateful to C\'ecile Monthus for her participation in the early stages of this work.
  This work has been
  supported by the Hungarian National Research Fund under grant No OTKA
K62588, K75324 and K77629. 
The Institut Jean Lamour is Unit\'e
  Mixte de Recherche CNRS No. 7198.
\end{acknowledgments}

\appendix
\section{Integral equations at the RF fixed point}
\label{app:1}
The probability distribution at this fixed point is a piece-wise function which satisfies different relations in the regions
$0<I<1$, $1<I<2$ and $I>2$, respectively. It has a delta peak at $I=0$ with strength $p_0$.

The strength of the delta peak renormalizes as
\begin{equation}
p_0'=P_0^2\;,
\label{p0}
\end{equation}
in terms of
\begin{equation}
P_0=2p_0-p_0^2+\int_{0}^{1}\!\!d I\, P(I) \int_0^{1-I} \!\!d x\, P(x)\;.
\label{P0}
\end{equation}
In the region $0<I<1$ we have the relation
\begin{equation}
P'(I)=2 P_0 P_1(I) + \int_0^I \!\!d x\, P_1(x) P_1(I-x)\;,
\label{P01}
\end{equation}
with
\begin{eqnarray}
P_1(I)&=&2 \int_{(I+1)/2}^1\!\! d x\, P(x) P(I+1-x) \cr
&+& 2 P(I) \int_{1}^{\infty}\!\! d x\, P(x),\quad 0<I<1\;.
\label{P_1}
\end{eqnarray}
In the region $1<I<2$ we have the relation:
\begin{eqnarray}
P'(I)&=&2 P_0 P_2(I) + \int_0^{I-1}\!\! d x\, P_1(x) P_2(I-x)\cr
&+&\int_{I-1}^1 \!\!d x\, P_1(x) P_1(I-x) \;,
\label{P12}
\end{eqnarray}
where the auxiliary function $P_2(I)$ is defined in Eq.~(\ref{P_2}).

Finally for $I>2$ the renormalization reads as:
\begin{eqnarray}
P'(I)&=&2 P_0 P_2(I) + \int_0^{1}\!\! d x\, P_1(x) P_2(I-x)\cr
&+&\int_{1}^{I-1} \!\!d x\, P_2(x) P_2(I-x) \;,
\label{P2}
\end{eqnarray}

\section{Integral equations at the Z fixed point}
\label{app:2}
The probability distribution at the Z fixed point is a piece-wise function which satisfies  different relations in the regions
$i<0$ and $i>0$, respectively. It has a delta peak at $i=0$, with strength $\pi_0$,
which renormalizes as
\begin{equation}
\pi_0'=\Pi_0^2\;,
\label{pi0}
\end{equation}
in terms of
\begin{equation}
\Pi_0=2p_0-p_0^2+\left[\int_{-\infty}^{0}\!\!d i \, \Pi(i) \right]^2\;.
\label{Pi0}
\end{equation}
In the region $i<0$ we have the relation
\begin{eqnarray}
\frac{1}{\alpha'}\Pi'(i)&=&2 \Pi_0 \Pi_1(i)+\int_{-\infty}^{i}\!\! d x\, \Pi_1(x) \Pi_2(i-x) \cr
&+&\int_i^0 \!\!d x\, \Pi_1(x) \Pi_1(i-x)
\label{Pim}
\end{eqnarray}
in terms of
\begin{equation}
\Pi_1(i)=2 \Pi(-i)\! \int_{-\infty}^{i}\!\!\!\! d x\, \Pi(x) + 2 \Pi(i)\! \int^{\infty}_{-i}\!\!\!\! d x\, \Pi(x),\quad i<0
\label{Pi1}
\end{equation}
and
\begin{equation}
\Pi_2(i)=2 \Pi(i) \int^{\infty}_{i} d x\, \Pi(x),\quad i>0\;.
\label{Pi2}
\end{equation}
Finally, in the region $i>0$, we have the relation
\begin{eqnarray}
\frac{1}{\alpha'}\Pi'(i)&=&2 \Pi_0 \Pi_2(i)+\int^{\infty}_{i}\!\! d x\, \Pi_2(x) \Pi_1(i-x) \cr
&+&\int^{i}_0\!\! d x\, \Pi_2(x) \Pi_2(i-x)\;.
\label{Pip}
\end{eqnarray}

\end{document}